\documentclass[prd,reprint]{revtex4-1}

\usepackage{lipsum}

\usepackage[utf8]{inputenc}

\usepackage{amsmath}
\usepackage{amssymb}
\usepackage{hyperref}

\begin{document}
 \title{Alien Calculus and non perturbative effects in Quantum Field Theory}
 \author{Marc P.~Bellon} 
\affiliation{
Sorbonne Universités, UPMC Univ Paris 06, UMR 7589, LPTHE, 75005, Paris, France}
\affiliation{CNRS, UMR 7589, LPTHE, 75005, Paris, France }

\begin{abstract}
In many domains of physics, methods are needed to deal with non-perturbative aspects. I want here to argue that a good approach is to work on the Borel transforms of the quantities of interest, the singularities of which give non-perturbative contributions.  These singularities in many cases can be largely determined by using the alien calculus developed by Jean \'Ecalle. My main example will be the two point function of a massless theory given as a solution of a renormalization group equation.
\end{abstract}

\maketitle

\section{Introduction}
At ECT\* in Trento, mathematicians and physicists met for the second time in September 2014 to exchange about their vision on Schwinger--Dyson equations. I, as a mathematical physicist, did not really know where to sit, but enjoyed the juxtaposition of concret problems and sophisticated tools. I was challenged to think a bigger picture for the use of a tool I recently learned, alien calculus. Here is an expanded and matured version of what I tried to explain to the participants.

Alien calculus is a technique initiated by Jean \'Ecalle to study functions which have non convergent Taylor expansions.  By providing tools for the study of the singularities of Borel transforms, it allows to compute or at least estimate the change the Borel sum incurs when the integration path crosses a line of singularities.  This sheds new lights for example on the Stokes phenomenon.  This allows also to give a new kind of expansions, called resurgent expansions, with explicit non-perturbative terms proportional to \(\exp(-f/a)\), which have all their derivatives at the origin vanishing. Singularities of the Borel transform on the positive axis, once viewed as liabilities introducing ambiguities in the summation process, become tools for a better understanding of the sum. The basic references are the works of \'Ecalle~\cite{Ecalle81}, but a gentler introduction can be found in~\cite{Sa14}.

The possibility of computing non perturbative terms is of tremendous interest in theories like QCD, since they are the only ones which can bridge physical and renormalization mass scales. In quantum mechanics, non perturbative contributions important for the computation of tunneling rates can be deduced from the action of solutions where the barrier is crossed in imaginary time, but the equivalent computation for a classically scale invariant theory as QCD is not clear, since the action of the classical instanton depends on the coupling constant: which value for the gauge coupling to choose, when the renormalization group makes it dependent on the scale? Alien calculus gives an alternative approach to determine where the singularities of the Borel transform are and what are their properties, and it does not present such ambiguities.

I will illustrate these general considerations by a study of the renormalization group equation for Borel transformed Green functions. This presentation is based on work done for the supersymmetric Wess--Zumino model~\cite{BeCl14}, which are presented in the contribution of Pierre Clavier in this volume. Our aim here is to show how general these methods can be and give hints on what results could be obtained in the future.

\renewcommand{\d}{\text{d}}
\section{Renormalization group}

It has been widely recognized that the renormalization group is a major feature of quantum field theory.  It corresponds to a breaking of the scale invariance in a classically scale invariant theory and is at the heart of the application of quantum field theory to critical phenomena in statistical mechanics as well as the application of QCD to the study of high energy hadronic interactions.  However, these applications are in fact limited to rather special cases of the renormalization group.  In statistical mechanics, one considers fixed points where scale invariance is recovered, eventually with anomalous dimensions, and for the applications to QCD and particle physics in general,  the scales where qualitatively different phenomena arise, e.g., the mass scale of hadrons, remain inaccessible.  However, if one wants to use the two-point functions defined through the renormalization group in some graphs to self-consistently compute renormalization group functions, one would need to get results for the entire kinematical space.

To see the nature of the problem, let us write the renormalization group for the ratio of a generic propagator to its free version \(G(a,L)\), depending on a coupling \(a\) and the logarithm of the invariant \(p^2\) of the momentum referred to a scale \(\mu^2\),  \(L = \log(p^2/\mu^2) \). This equation involves the anomalous dimension \(\gamma\) and the scale dependence of the coupling \(\beta\):
\begin{equation}
\label{RGs}
			\frac{\partial}{\partial L} G(a,L) = (\gamma(a) + \beta(a) a \partial_a) G(a,L).
\end{equation}
It is easy to use such an equation to obtain the expansion of \(G\) with respect to \(L\) to any finite degree, but higher degree terms will involve higher derivatives of the functions \(\gamma\) and \(\beta\), except in the case of a fixed point \(a_0\) where \(\beta(a_0)\) vanishes or in the extended supersymmetry cases where the renormalization group functions reduce to their lower order part. In more realistic theories, only a few terms of the perturbative expansions of the renormalization group functions are known, we can presume that these perturbative expansions are always divergent  and having the functional dependence of \(G\) on \(L\) seems out of reach.  

\section{Borel plane formulation}

A solution is nevertheless possible if we consider Borel transforms of all these functions.  Indeed, it is desirable to consider the Borel transforms, which are probably well defined as analytic functions in the vicinity of the origin, instead of the purely formal original power series. At the level of power series, the Borel transform is defined by:
\begin{equation}
\label{Borel}
f(a) = \sum_{n=0}^\infty f_n a^{n+1} \quad \longrightarrow \quad \hat f(\xi) = \sum_{n=0}^\infty \frac {f_n}{n!} \xi^n.
\end{equation}
The Borel transformation transforms pointwise products of functions into convolution products and equations like the renormalization group equation (\ref{RGs}) can be translated to an equation for the Borel transforms. In its simplest version, the Borel transform involves a shift of degree, so that we isolate the constant part in \(G\), so that \(\hat G\) is in fact the Borel transform of \(G -1 \) viewed as a function of the variable \(a\) and the degree counting operator \(a \partial_a\) becomes \(\partial_\xi \xi\), which we write as \(\nabla \Xi\), with \(\Xi\) the operator of multiplication by the variable in the Borel plane.
Noting \(\hat G(\cdot,L)\) for the function evaluating to \(\hat G(\xi,L)\) at \(\xi\), equation~(\ref{RGs}) becomes
\begin{equation}
\label{RGb1}
	\frac{\partial}{\partial L} \hat G(\cdot,L) = \hat\gamma + \hat\gamma \star \hat G(\cdot,L)
		+ \hat\beta \star \nabla \Xi\; \hat G(\cdot,L)
\end{equation}
At this point, one may think that nothing changed, since there still is a derivative acting on \(\hat G\). However the convolution product is defined through an integral on which an integration by part transformation can be performed.
\begin{equation*}
  f \star g ( \xi ) = \int_0^\xi f(\eta) g(\xi-\eta) d\eta
\end{equation*}
so that 
\begin{equation}
f\star g'   = f' \star g + f(0) g - g(0) f.
\end{equation}
If we denote by \(\beta_0\) and \(\beta_1\) the first two coefficients of the \(\beta\) function, so that 
\begin{equation*}
  \beta(a) = \beta_0 a + \beta_1 a^2 + \cdots, \quad\quad \hat\beta(\xi) = \beta_0 + \beta_1 \xi + \cdots,
\end{equation*}
the renormalization group equation in the Borel plane~(\ref{RGb1}) becomes
\begin{equation}
\label{RGb2}
	\left( \frac{\partial}{\partial L} - \beta_0 \Xi \right) \hat G(\cdot,L) = \hat\gamma  + \hat\gamma \star  \hat G(\cdot,L) + \hat\beta' \star \Xi \hat G(\cdot,L).
\end{equation}
In the first form of the renormalization group equation~(\ref{RGb1}), treating the convolution products as perturbations gives a solution for \(\hat G\) as a power series in \(L\), but now, the \(\beta_0\Xi\) term gives an exponential as the order 0 term. Using the renormalization condition that \(G(a,0) =1 \), or equivalently, \(\hat G(\xi,0) =0\), we get for the order 0 solution
\begin{equation}
\label{order0}
	\hat G_0(\xi,L) =\frac{ \hat \gamma(\xi) }{\beta_0 \xi} \bigl( e^{\beta_0 \xi L} - 1\bigr).
\end{equation}
Introducing this order zero solution in the convolution products of the renormalization group equation~(\ref{RGb2}) will give successive approximations which will have  in common to be superpositions of those exponentials. The proper parameterization of this superposition is however non trivial, since there are singular contributions from the end points of the interval \(0\) and~\(\xi\).  Convolution products can also be defined by integrating over a path different from the straight line and this is important to prevent spurious singularities to appear when one defines analytic extensions. It is therefore not clear how to obtain an efficient representation for \(\hat G\). Nevertheless, \(e^{\beta_0 \xi L}\) is \(p^2/\mu^2\) to the power \(\beta_0 \xi\). This means that larger powers of \(p^2\) will appear in the representation of the propagator the further we go from the origin. Potentially, this means that all sorts of new divergences could appear when looking for the analytic extension of \(\hat G\) which is necessary to go back to the propagator \(G\) in terms of the coupling \(a\).  The nature of the divergences depends fundamentally on the sign of \(\beta_0\), making a clear distinction between asymptotically free theories and the other ones: in asymptotically free theories with \(\beta_0\) negative, we obtain negative powers of \(p^2\) for the positive values of \(\xi\) necessary to evaluate the propagator for physical values of the coupling \(a\) and so we do not have new ultraviolet  divergences, but infrared ones. The situation is reversed if \(\beta_0\) is positive.

\section{Parameterization of the propagator}

Nevertheless a good parameterization of the propagator is  possible as a contour integral~\cite{BeCl14}: 
\begin{equation} \label{param_G}
 \hat{G}(\xi,L) = \oint_{\mathcal{C}_{\xi}} f(\xi,\zeta) e^{\zeta L} \frac{\d\zeta}{\zeta}
\end{equation}
with $\mathcal{C}_{\xi}$ any contour enclosing $0$ and $\beta_0 \xi$. On a contour minimally including the endpoints, the jump of \(f\) along a cut from \(0\) to \(\beta_0 \xi\) gives a smooth integral, while the singularities at the end points will contribute singular terms.  The condition that \(\hat G(\xi,0)\) is zero is also easily obtained in this formalism, since the exponential becomes 1 for \(L=0\) and the contour can be expanded to infinity.  It is therefore sufficient that \(f\) have limit 0 at infinity. The renormalization group equation for $\hat{G}$~(\ref{RGb2}) becomes an equation on $f$, since one can use the same contour for the computation of \(\hat G\) for all the necessary values of \(\eta\) and then, switching the order of the contour integral and the other operations, one can write everything 
as a contour integral on a common path, since even the \(L\) independent term can be given the form of a similar contour integral, using that
\begin{equation*}
 1 = \oint_{\mathcal{C}_{\xi}}e^{\zeta L}\frac{\d\zeta}{\zeta}.
\end{equation*}
The definition of the contour integral \(\oint\) includes a factor \(1/(2\pi i)\) to simplify notations. The annulation of the integrand is certainly a sufficient condition for the renormalization group equation~(\ref{RGb2}). One ends up with the following equation for the functions $f_\zeta$ defined by \(f_\zeta(\xi) = f(\xi,\zeta)\):
\begin{equation} \label{renorm_f}
 (\zeta-\beta_0 \Xi) f_\zeta = \hat\gamma + \hat\gamma \star f_\zeta + \hat\beta' \star \Xi f_\zeta
 \end{equation}
 We can expand this equation in powers of \(\zeta^{-1}\) and obtain an expansion which exactly corresponds to the expansion of \(G\) in powers of \(L\) by evaluating the contour integral by its residue at infinity. This however would defeat the purpose of this representation of the propagator which was precisely to avoid an expansion in power of \(L\). This however shows that equation~(\ref{renorm_f}) is necessary, since as a function of \(\zeta\), it is holomorphic in a neighborhood of \(\infty\) and has all of its Taylor coefficients 0.

Exchanging again the order of integrations, the evaluation of the graphs appearing in Schwinger--Dyson equations with the modified propagators amount to first evaluate them with some propagators modified with the powers \(\zeta\) of the momenta, the Mellin transform of the graph. The perturbative solution of the Schwinger--Dyson equations only use the derivatives of this Mellin transform at the origin, but analytic continuation of the Borel transform depends on its value at large. For small enough \(\zeta\), graphs will depend holomorphically on \(\zeta\): this is known since a long time because such modifications of the propagators are at the heart of analytical regularization~\cite{Sp69,Sp71}, which had its limelight moment before being supplanted by dimensional regularization. Indeed, holomorphy in the domain where the diagram is convergent is quite natural since the integrand depends holomorphically on the \(\zeta\)'s. However, for larger values of \(\zeta\), new divergences can appear but the Mellin transform can be shown to have a meromorphic extension with simple poles~\cite{BeSc12}, in particular, infrared poles will appear for any negative integer value for \(\zeta\). We therefore keep the independence on the precise form of the integration contour that is important for the validity of the representation~(\ref{param_G}), but when shifting the integration contour \(\mathcal{C}_\xi\) to accommodate a larger value of \(\xi\), the crossing of these poles adds terms proportional to \(f_{-k}\) which will contribute to the \(\hat\gamma\) and \(\hat\beta\) functions. 

\section{Alien calculus and the singularities of the Borel transform}
The alien calculus introduced by \'Ecalle allows us to show that  \(f_{\zeta}\) has a singularity for \(\xi = \zeta/\beta_0\), with a computable exponent. Indeed, there is an operator \(\Delta_{\xi_0}\) which allows to extract the singular part of a function around the point \(\xi_0\) and becomes a derivation with respect to the convolution product if one takes certain weighted averages of the analytic continuations along the different classes of paths. Applying this operator \(\Delta_{\xi_0}\), called an alien derivative by \'Ecalle, to a system of equations satisfied by the propagator and the renormalization group functions therefore produces a linear system of equations for the alien derivatives of these functions.  For generic values of \(\xi_0\), the only possible solution for this system of equations is zero. This means that \(\xi_0\) cannot be the position of a singularity of the solution unless \(\xi_0\) is the sum of the positions of other singularities\footnote{Convolution products have generically singularities at the positions of the singularities of each factors, but also at the sum of the positions, at least in some of the sheets of the Riemann surface on which the convolution product is defined.}. 

In the case of equation~(\ref{renorm_f}), the right hand side, which involves convolution products, will become less singular than \(\Delta_{\xi_0} f_\zeta\), so that a singularity is possible only if \(\zeta- \beta_0\xi_0\) vanishes to allow the singularity degree of both sides to match. For generic values of \(\zeta\), such a singularity has no particular effects on the solution of the theory, since one can always shift the integration contours to find an expression of the results which does not depend on this particular \(f_\zeta\).  However the contributions of the residues of the poles which must be crossed when extending the integration contour will be proportional to \(f_{-k}\) for some integer \(k\) in the case of infrared poles. In this way,  singularities will therefore also appear in the Borel transforms of the renormalization group functions and the propagator for \(\xi = -k/\beta_0\). If \(\beta_0\) is negative, these singularities will be for positive real values of \(\xi\) and will allow us to compute non-perturbative contributions to these quantities, proportional to \(e^{1/\beta_0 a}\) and its powers. Indeed, in Ecalle's theory of summation, links exist between alien calculus and transseries, through the freedom in the definition in transmonomials when only their values on a small sector are considered. The whole procedure of resummation has been detailed in the book~\cite{Ec92}, but exploiting it would largely outstrip this elementary presentation.

Going from this rapid sketch to a fully consistent evaluation of these effects will not be simple, but we are confident that an analytic approach to non-perturbative phenomena in quantum field theories can be obtained along these lines and provide new insights on such questions as the confinement of color and the generation of mass in QCD, fulfilling the intuitions of Manfred Stingl~\cite{St02}. The singularities at \(-k/\beta_0\) should therefore translate in terms proportional to \((\Lambda^2_{\mathrm{QCD}}/p^2 )^k\) which look like very bad news for the infrared behavior of the propagator.  However, such singularities have contribution from \(\Delta_{-k/\beta_0}\) but also from the effect of all products of alien derivatives such that the sum of the arguments is \(-k/\beta_0\), in particular the \(k\)-th power of \(\Delta_{-1/\beta_0}\). If this last term is dominant, one should be able to see all these terms as the expansion of a massive propagator, with a squared mass proportional to \(\Lambda^2_{\mathrm{QCD}}\). To control the infrared behavior of our theory, we have gone to a representation with even worse infrared behavior but which can be seen as the badly behaving expansion of a perfectly fine theory.  However, the bad infrared behavior is unavoidable for the Borel transform which is a necessary step in the resolution of this conundrum since infrared divergences feed the singularities of the Borel transform which produce these non-perturbative terms which finally can be interpreted as mass terms.

Since the mass terms should all be linked to the same alien derivative of the \(\beta\)-function, ratio of masses of hadrons should be computable in the chiral limit of vanishing quark masses. The link with the fundamental scale \(\Lambda^2_{\mathrm{QCD}}\) would be harder to get since it requires comparing the actual terms in the perturbative coefficients of the \(\beta\)-function and the asymptotic behavior that can be deduced from the singularities of the Borel transform. As in~\cite{BeCl13}, some aspects of the computation of the possible forms of the alien derivatives could be easier when using a formal multiplicative model of transseries, but the final interpretation is better seen at the level of the Borel transform where we deal with actual functions.
\bibliography{renorm}
  \end{document}